\documentclass[10pt,journal,letterpaper]{IEEEtran}
\usepackage[T1]{fontenc}
\usepackage{amsmath,amssymb}
\usepackage{graphicx}
\usepackage{booktabs,makecell}
\usepackage{cite}
\usepackage{microtype}
\usepackage{balance}
\usepackage[hidelinks]{hyperref}
\hypersetup{
  pdftitle={qVocalTract: Efficient Encoding Framework of Vocal-Tract Spectra on a Quantum Computer},
  pdfauthor={Sheng Li and Jiyi Li}
}
\begin{document}

\title{qVocalTract: Efficient Encoding Framework of Vocal-Tract Spectra on a Quantum Computer}
\author{Sheng~Li and Jiyi~Li%
\thanks{Sheng Li is with the Graduate School of Informatics, Kyoto University, Kyoto 606-8501, Japan; the School of Engineering, Institute of Science Tokyo, Yokohama, Kanagawa 226-8501, Japan; and TRIP, RIKEN, Wako, Saitama 351-0198, Japan (corresponding author; e-mail: sheng.li@ieee.org).}%
\thanks{Jiyi Li is with the Graduate School of Information Science and Technology, Hokkaido University, Sapporo, Hokkaido 060-0814, Japan (e-mail: jyli@ist.hokudai.ac.jp).}}
\maketitle

\begin{abstract}
The vocal-tract transfer function links the physical articulatory model to speech. Quantum encoding of vocal-tract transfer functions requires circuits that preserve acoustic structure under hardware noise. This paper proposes the qVocalTract framework, which uses optimized, ancilla-free staircase layers for spectral loading. Simulations and a superconducting processor compare this approach with sequential matrix-product-state preparation on vocal-tract data. Staircase preparation lowers spectral error and improves resonance recovery, while a modal probe demonstrates coherent processing. The comparison reveals a tradeoff between approximation accuracy and accumulated circuit noise. These findings guide practical support for further quantum acoustic processing with physical vocal-tract models.
\end{abstract}


\section{Introduction}
Vocal-tract transfer functions link articulatory geometry to acoustic resonances. They provide a physical test case for loading structured acoustic data onto quantum processors. Amplitude encoding stores $N=2^n$ values in $n$ qubits. However, generic state preparation requires a gate count that grows with $N$~\cite{Plesch2011,Shende2006}. The loading circuit is therefore central to a practical quantum representation.

Li et al.~\cite{LiLi2026} studied matrix-product-state (MPS) compression of transfer functions from the Dresden Vocal Tract Dataset (DVTD)~\cite{Birkholz2020}. An MPS represents a state with local tensors connected by a bond index~\cite{Vidal2003,Oseledets2011}. Sequential preparation carries this index through an ancilla register~\cite{Schon2005}. A larger bond dimension improves approximation capacity. It also enlarges each site unitary and increases ancilla-routing costs on a sparsely connected processor.

We implement qVocalTract on \texttt{ibm\_kawasaki} by adapting ancilla-free staircase decompositions and local unitary optimization to acoustic spectra~\cite{Ran2020,Rudolph2024,BenDov2024}. Our contribution is a hardware comparison that connects circuit cost to spectral-envelope error and resonance recovery. We evaluate a reference vowel and all 44 DVTD cases, then test coherent evolution with a separate modal probe. Figure~\ref{fig:circuits} summarizes the workflow and preparation circuits.

\section{Methods}
\subsection{Encoding and sequential preparation}
We sample a calculated transfer function $G(f)$ at $f_m=m f_{\max}/N$, with $N=2^{14}$ and $f_{\max}=10$\,kHz. The grid includes zero and excludes $f_{\max}$. The peak-normalized log-magnitude and encoded state are
\begin{equation}
\begin{aligned}
x_m &= 20\log_{10}|G(f_m)|-\max_j20\log_{10}|G(f_j)|,\\
|\psi\rangle &= \|x\|_2^{-1}\sum_{m=0}^{N-1}x_m|m\rangle.
\end{aligned}
\label{eq:state}
\end{equation}
All target amplitudes are nonpositive. Qubit $i$ carries bit $i$ of $m$. Right-canonical tensor-train singular-value decomposition (SVD), truncated at bond dimension $\chi$, produces a normalized approximation $|\psi_\chi\rangle$. For 14 sites, the stored tensor counts are 104, 360, and 1192 for $\chi=2,4,8$.

\begin{figure*}[!t]
\centering
\includegraphics[width=\textwidth]{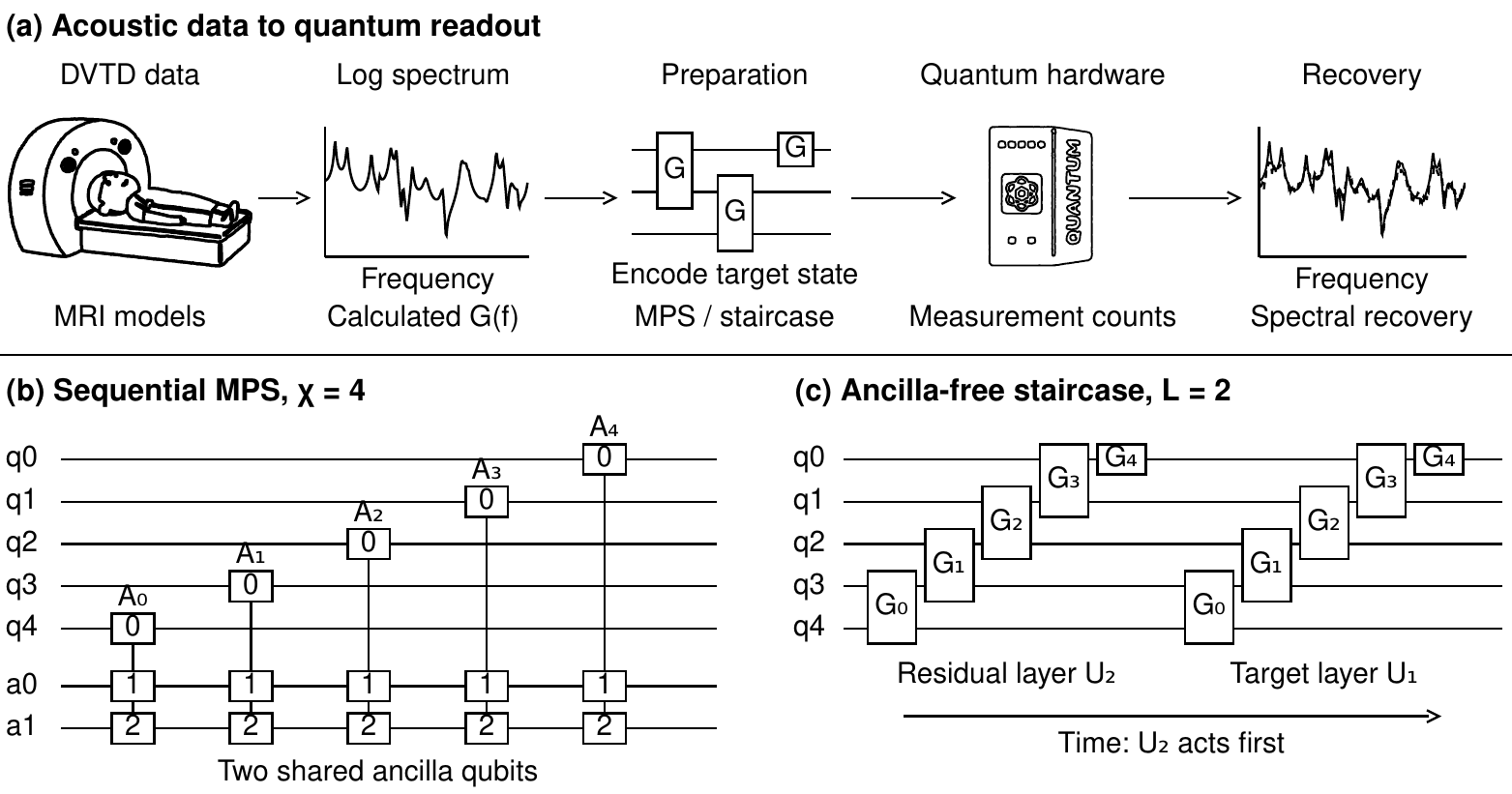}
\caption{(a)~Workflow from DVTD transfer functions to quantum readout. Curves illustrate the reference-vowel target (solid) and raw three-layer reconstruction (dashed). (b)~Sequential MPS preparation~\cite{Schon2005} for five data qubits: each site unitary shares the ancilla register; numerals label its ports. (c)~Two staircase layers without ancillas; time runs left to right.}
\label{fig:circuits}
\end{figure*}

Sequential preparation extends each site isometry to a unitary on one data qubit and $\lceil\log_2\chi\rceil$ ancilla qubits. The ancilla carries the bond between sites and returns to zero at the end [Fig.~\ref{fig:circuits}(b)]. Larger site unitaries require more two-qubit gates~\cite{Shende2006}. The shared ancilla may also need routing between distant data qubits. We report compiled two-qubit gate counts.

\subsection{Ancilla-free staircase preparation}
A bond-two MPS requires $n-1$ nearest-neighbor two-qubit gates and one single-qubit gate~\cite{Ran2020}. The next data qubit carries the bond, removing the separate ancilla [Fig.~\ref{fig:circuits}(c)]. We construct the first staircase $U_1$ from the target and the next from the residual state $U_1^\dagger|\psi\rangle$. Repeating this step gives $U_1U_2\cdots U_L|0\rangle$. The last constructed layer is applied first.

For $L\ge2$, we refine the gates with local polar-decomposition updates. Write the target overlap for one gate $g$ as $\mathrm{Tr}(gE)$. An SVD $E=W\Sigma V^\dagger$ gives the maximizing unitary $g=VW^\dagger$. Optimization runs for at most 40 forward sweeps and stops when the fidelity gain falls below $10^{-7}$. Fidelity is $|\langle\psi|\phi\rangle|^2$. The one-layer baseline uses the unoptimized bond-two staircase. Gates connect adjacent logical qubits; the physical layout and transpiler determine routing.

\subsection{Envelope and resonance readout}
Computational-basis measurements estimate $P(m)=|\phi_m|^2$ for the prepared state. Using the known target sign and scale gives $\hat x_m=-\|x\|_2\sqrt{\hat P(m)}$. This readout estimates amplitude magnitudes; it does not verify relative signs or phases in an approximate state. Since 50\,000 shots provide only about three samples per full-resolution outcome, we combine $B=64$ adjacent bins:
\begin{equation}
\hat x_B[b]=-\|x\|_2\sqrt{\frac{1}{B}\sum_{m\in b}\hat P(m)}.
\label{eq:readout}
\end{equation}
The 256-bin envelope has a spacing of 39.06\,Hz. We compute RMSE over 0--10\,kHz against the target reduced by block RMS at the same resolution. Resonances are the first three peaks between 150 and 4000\,Hz, with at least 3\,dB prominence and about 200\,Hz separation. Peak detection uses no smoothing for the reference vowel and a three-bin moving average for the dataset sweep. These settings reproduce the respective archived records. Missing peaks count as failures for the all-three-within-100-Hz criterion. Formant MAE uses available target--reconstruction pairs and is reported with the success fraction.

We also reproduce the recorded tensor-product readout correction~\cite{Bravyi2021,Nation2021}. It uses all-zero and all-one calibration circuits to correct the coarse marginal and assumes independent readout errors. We report raw-count comparisons alongside the corrected results.

\section{Experimental settings}
DVTD contains 44 vocal-tract configurations from two speakers, including vowels and consonants~\cite{Birkholz2020}. Calculated transfer functions define the targets. Measured transfer functions provide an additional acoustic reference. We interpolate magnitude and unwrapped phase separately. Both envelopes use the calculated peak as their dB offset. We evaluate block-RMS error over the full band; the earlier study independently normalized the envelopes and evaluated 80--5000\,Hz.

We conducted hardware experiments on \texttt{ibm\_kawasaki}. We used Qiskit~\cite{Qiskit2024}, transpiler optimization level 3, fixed seeds, and XY4 dynamical decoupling on hardware~\cite{Ezzell2023}. We compare hardware with an ideal sampler and the static \texttt{FakeTorino} model, which represents a different device calibration. The reference case is s1-01, tense /a/ from speaker 1. Its target resonances are 527, 1113, and 2285\,Hz. The spectrum experiments use 100\,000 shots for the ideal sampler, 20\,000 for the noise model, and 50\,000 for hardware. The dataset sweep uses 20\,000 shots per case. Sampling budgets, therefore, differ in the single-case noise comparison.

\section{Results}
\subsection{Reference vowel}
\begin{figure*}[!t]
\centering
\includegraphics[width=\textwidth]{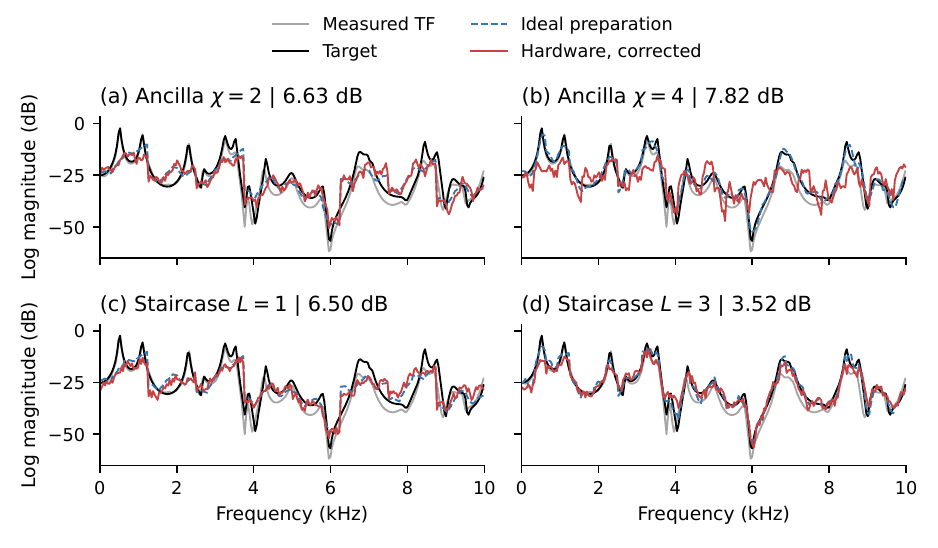}
\caption{Reference-vowel envelopes reconstructed from the archived hardware counts at 256-bin resolution. Panels show sequential $\chi=2,4$ and staircase $L=1,3$ preparations; titles give corrected RMSE against the target. All curves and Table~\ref{tab:single} use the same block-RMS convention.}
\label{fig:spectrum}
\end{figure*}

Table~\ref{tab:single} separates approximation, ideal-sampler, and hardware errors. Increasing $\chi$ improves noiseless sequential preparation. On hardware, however, RMSE rises from 6.63 to 10.30\,dB as the compiled gate count grows from 39 to 1839. Three staircase layers use 98 gates and achieve 3.52\,dB. Their resonances are 488, 1113, and 2246\,Hz, each within one coarse bin of the target. Four layers improve the ideal approximation but raise hardware error to 4.88\,dB. Figure~\ref{fig:spectrum} shows selected envelopes at the same 256-bin resolution.

\begin{table*}[!t]
\caption{Reference vowel on \texttt{ibm\_kawasaki}: 256-bin envelope RMSE (dB), including the approximation column, and detected resonances (Hz). Target resonances are 527, 1113, and 2285\,Hz. ``2q'' denotes the compiled CZ count. Recovered values use the recorded readout correction.}
\label{tab:single}
\centering
\setlength{\tabcolsep}{3pt}
\small
\renewcommand{\arraystretch}{1.1}
\begin{tabular*}{0.92\textwidth}{@{\extracolsep{\fill}}lrrrrl@{}}
\toprule
Preparation & 2q & Approx. & Ideal & Device & \makecell{Resonances\\(Hz)} \\
\midrule
Ancilla $\chi=2$ & 39 & 6.22 & 6.35 & 6.63 & 879, 1152, 2011 \\
Ancilla $\chi=4$ & 461 & 2.61 & 2.60 & 7.82 & 410, 722, 1074 \\
Ancilla $\chi=8$ & 1839 & 0.26 & 0.71 & 10.30 & 722, 918, 1465 \\
qVocalTract $L=1$ & 29 & 6.22 & 6.26 & 6.50 & 722, 1113, 1308 \\
qVocalTract $L=2$ & 67 & 4.11 & 4.13 & 4.51 & 879, 1777, 2285 \\
qVocalTract $L=3$ & 98 & 2.32 & 2.39 & 3.52 & 488, 1113, 2246 \\
qVocalTract $L=4$ & 130 & 1.89 & 2.03 & 4.88 & 449, 761, 1074 \\
\bottomrule
\end{tabular*}
\end{table*}

\subsection{Dataset-wide comparison}
\begin{figure*}[!t]
\centering
\includegraphics[width=\textwidth]{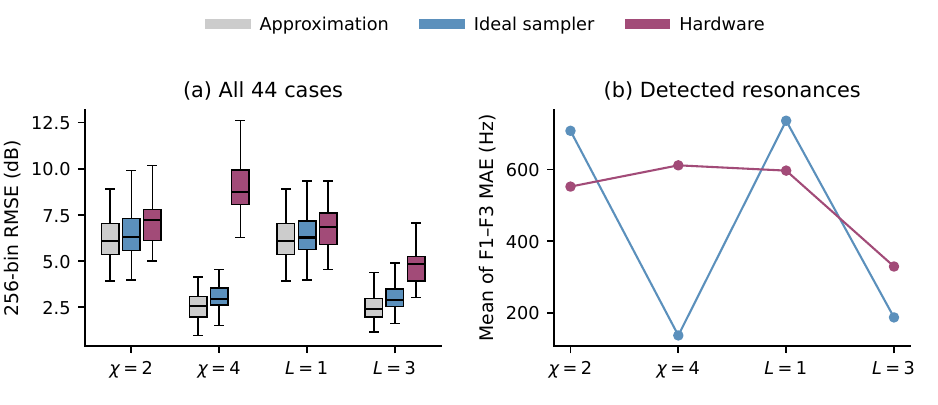}
\caption{44-case sweep with the recorded readout correction. Left: 256-bin RMSE per preparation; grey = noiseless approximation, blue = ideal sampler, purple = \texttt{ibm\_kawasaki}. Right: mean formant error on the ideal sampler and on the device.}
\label{fig:sweep}
\end{figure*}

Across 44 cases, the three-layer staircase gives a corrected mean hardware RMSE of 4.74\,dB. Sequential $\chi=2$ and 4 give 7.15 and 9.15\,dB, respectively (Table~\ref{tab:sweep}; Fig.~\ref{fig:sweep}). Three layers improve on both baselines in every case. The fraction with all three resonances within 100\,Hz is 0.30 for three layers and zero for both sequential baselines. Resonance recovery improves, although errors in the second and third peaks remain substantial.

Raw-count recalculation gives mean RMSEs of 4.77, 7.16, and 9.22\,dB for three layers and sequential $\chi=2,4$. The staircase again wins all 44 comparisons against each baseline. The same 13 of 44 cases meet the three-resonance criterion before and after correction. The improvement, therefore, persists without readout correction. Raw three-layer resonance MAEs are 157, 324, and 522\,Hz, based on 44, 44, and 43 valid pairs. One target has no detected third peak.

Against measured transfer functions, the three-layer hardware RMSE is 5.96\,dB, compared with 4.75\,dB for its noiseless approximation. Sequential $\chi=8$ was evaluated only for the reference vowel and was excluded from the 44-case sweep.

\begin{table*}[!t]
\caption{All 44 cases on \texttt{ibm\_kawasaki}, with recorded readout correction: mean 256-bin RMSE (dB), MAE of detected resonances (Hz), and the fraction with all three within 100\,Hz (``ok''). ``2q'' is the mean compiled gate count rounded to the nearest integer; 20\,000 shots per circuit.}
\label{tab:sweep}
\centering
\setlength{\tabcolsep}{3pt}
\small
\renewcommand{\arraystretch}{1.1}
\begin{tabular*}{0.92\textwidth}{@{\extracolsep{\fill}}lrrrrlr@{}}
\toprule
Preparation & 2q & Approx. & Ideal & Device & \makecell{MAE F1/F2/F3\\(Hz)} & ok \\
\midrule
Ancilla $\chi=2$ & 39 & 6.26 & 6.50 & 7.15 & 452 / 586 / 619 & 0.00 \\
Ancilla $\chi=4$ & 461 & 2.56 & 3.06 & 9.15 & 222 / 563 / 1050 & 0.00 \\
qVocalTract $L=1$ & 30 & 6.26 & 6.50 & 6.88 & 597 / 570 / 624 & 0.02 \\
qVocalTract $L=3$ & 99 & 2.48 & 2.99 & 4.74 & 160 / 310 / 518 & 0.30 \\
\bottomrule
\end{tabular*}
\end{table*}

\subsection{Noise-model comparison}
Table~\ref{tab:fake} reports the reference-vowel results under \texttt{FakeTorino}. Sequential circuits degrade as $\chi$ increases under both modeled and hardware noise. The model slightly favors four layers (4.21 versus 4.27\,dB), while hardware favors three (3.52 versus 4.88\,dB). This difference shows how circuit noise affects the best preparation depth.

\begin{table}[!t]
\caption{Reference-vowel 256-bin RMSE (dB) on the ideal sampler, static \texttt{FakeTorino} noise model, and hardware. Parentheses give the gate count compiled for the noise model.}
\label{tab:fake}
\centering
\setlength{\tabcolsep}{3pt}
\begin{tabular}{lrrr}
\toprule
Preparation (2q) & Ideal & Noise model & Device \\
\midrule
Ancilla $\chi=2$ (39) & 6.35 & 6.42 & 6.63 \\
Ancilla $\chi=4$ (452) & 2.60 & 6.51 & 7.82 \\
Ancilla $\chi=8$ (1850) & 0.71 & 8.91 & 10.30 \\
qVocalTract $L=1$ (29) & 6.26 & 6.34 & 6.50 \\
qVocalTract $L=2$ (67) & 4.13 & 4.92 & 4.51 \\
qVocalTract $L=3$ (98) & 2.39 & 4.27 & 3.52 \\
qVocalTract $L=4$ (130) & 2.03 & 4.21 & 4.88 \\
\bottomrule
\end{tabular}
\end{table}

\subsection{Separate modal-probe experiment}
\begin{figure*}[!t]
\centering
\includegraphics[width=\textwidth]{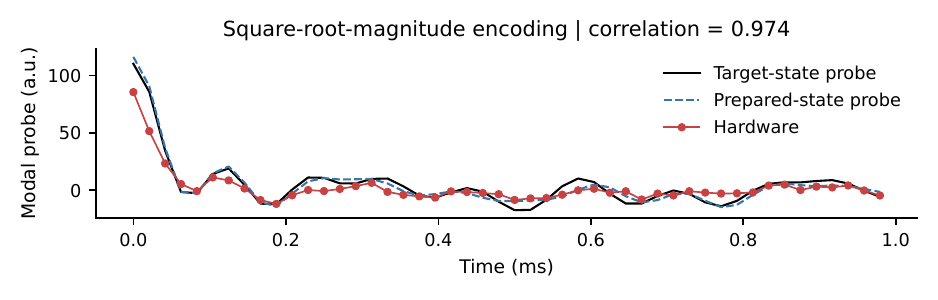}
\caption{Modal probe with square-root-magnitude encoding and a one-layer staircase: 48 samples at 48\,kHz on \texttt{ibm\_kawasaki}. Curves show the target-state, prepared-state, and hardware probes.}
\label{fig:waveform}
\end{figure*}

The modal-probe experiment uses square-root-magnitude encoding rather than the log-magnitude state in Eq.~\eqref{eq:state}. Its real amplitudes satisfy $\eta_m\propto\sqrt{|G(f_m)|}$ and $\sum_m\eta_m^2=1$ (the code's \texttt{cuda} encoding). For real prepared amplitudes $\phi_m$, let $S_0=\sum_m\phi_m\ne0$. The modal probe is $y(t)=\mathrm{Re}\sum_m\phi_m e^{-\mathrm{i}2\pi f_m t}$. Linear frequency spacing implements diagonal evolution with single-qubit phase gates. For interference readout, we add a clock qubit in $|+\rangle$ and use it to control the evolution. Hadamard gates on the clock and data register yield joint probabilities satisfying
\begin{equation}
y(t)=\frac{N}{S_0}\{P(c=0,m=0)-P(c=1,m=0)\}.
\end{equation}
Controlled phase gates add entangling operations after preparation. With one staircase layer, the hardware experiment uses up to 84 two-qubit gates per circuit, 48 time samples at 48\,kHz, and 2000 shots per sample. Correlation with the ideal prepared-state probe is 0.974 (Fig.~\ref{fig:waveform}). The noise-model and ideal-sampler correlations are 0.989 and 0.998. This experiment tests coherent processing of the prepared state, rather than speech synthesis.

\section{Discussion}
The results connect state-preparation cost to acoustic accuracy. Larger sequential bond dimensions reduce approximation error but increase decomposition and routing costs. Staircase layers add local two-qubit gates in smaller steps. This gives a useful balance between approximation accuracy and accumulated noise on the tested processor.

Classical optimization uses the full target vector for each case, and envelope recovery requires many samples. Peak extraction is sensitive to missing and spurious resonances, so repeated hardware runs are needed to estimate variability. 

\section{Conclusion}
In this paper, the ancilla-free staircase preparation improves spectral recovery over the tested sequential baselines. Three layers provide the best hardware accuracy for the reference vowel and improve mean accuracy across 44 configurations. Acoustic error and resonance recovery provide practical criteria for choosing preparation circuits. Extending this approach to broader physical vocal-tract models and coherent acoustic operations is the next step toward quantum speech processing.

\section*{Acknowledgment}
This work was supported in part by a research grant from The Telecommunications Advancement Foundation and by JSPS KAKENHI Grant Number JP23K11227.

\balance
\bibliographystyle{IEEEtran}
\bibliography{refs}
\end{document}